\documentstyle{article}
\newtheorem{Theorem} {Theorem} [section]

\newtheorem{Proposition} [Theorem] {Proposition}
\newtheorem{Conjecture}[Theorem]{Conjecture}
\newtheorem{Problem}[Theorem]{Problem}
\newtheorem{Definition}[Theorem]{Definition}

{\catcode `\@=11 \global\let\AddToReset=\@addtoreset}
\AddToReset{equation}{section}

\newcommand{\subjclass}[1]{}

\newcommand{\R}{I\!\! R}

\newcommand{\bysame}{---}
\newcommand{\proof}{{\sc Proof:}\ }
\def\scri{\hbox{${\cal J}$\kern -.645em {\raise
      .57ex\hbox{$\scriptscriptstyle (\ $}}}}
\newcommand{\Scri}{\scri} \newcommand{\cB}{{\cal B}}
\newcommand{\cW}{{\cal W}}
\newcommand{\cD}{{\cal D}}
\newcommand{\cDs}{{\cal D}(\Sigma)}
\newcommand{\cC}{{\cal C}}
\newcommand{\prt}{\partial}
\newcommand{\rmint}{\mbox{\footnotesize\rm int}}
\newcommand{\veps}{\Sigma} \newcommand{\eq}[1]{(\ref{#1})}
 \newcommand{\D}{{\cal D}}
\newcommand{\cU}{{\cal U}}
\newcommand{\cN}{{{\cal N}}{}}
\newcommand{\cNX}{{{\cal N}[X]}{}}
\newcommand{\cSX}{{{\cal S}[X]}{}}

\newcommand{\commentout}[1]{}

\newcommand{\ee}{\end{equation}} \newcommand{\bea}{\begin{eqnarray}}
\newcommand{\eea}{\end{eqnarray}}
\newcommand{\beaa}{\begin{eqnarray*}}
\newcommand{\eeaa}{\end{eqnarray*}}

\begin{document}

\title{``No Hair" Theorems -- Folklore,   Conjectures, Results}
\author{Piotr T.\ Chru\'sciel}
 \date{gr-qc/9402032}

\maketitle

\begin{abstract}
  Various assumptions underlying the uniqueness theorems for black
holes are discussed. Some new results are described, and various
unsatisfactory features of the present theory are stressed.
\end{abstract}
\section{Folklore, Conjectures}
\label{introduction}
A classical result in the theory of black holes, known under the name
of ``no--hair Theorem'', is the following:
\begin{Theorem}
\label{T0}
Let $(M,g)$ be a good electro--vacuum  space--time with a non--empty
black hole region and with a Killing vector
field which is timelike in the asymptotic regions. Then $(M,g)$ is
diffeomorphically isometric to a Kerr--Newman space--time or to a
Majumdar--Papapetrou space--time.  \end{Theorem}

This Theorem is known to be true under various definitions of ``good
space--time'' (all of which actually imply that it {\em cannot} be a
Majumdar--Papapetrou space--time), and the purpose of this paper is to
discuss various problems related to the as--of--today--definition of
``good space--time'' needed above. Clearly, one would like to have a
definiton of ``good space--time'' as weak as possible. Moreover one
would like this definition to have some degree of verifiability and
``controllability'', and to be compatible with our knowledge of the
structure of the theory gained by some perhaps completely different
investigations.

We shall focus here on uniqueness theory of stationary electro--vacuum
space--times. It is, however, worthwile mentioning that substantial
progress has been made recently in the understanding of various other
models. In particular one should mention various results about
uniqueness of perfect fluid models \cite{MuAL,BS}, $\sigma$--models
\cite{MH1,MH2,HS}, Einstein--Yang--Mills solutions
\cite{BP,Bizonwlochowatosc,SW1,SW2}, and
 dilatonic black holes \cite{MuAdilaton}, {\em cf.\/} also
 \cite{Bizonwlochowatosc,Gibbons}. Many of the questions
raised here as well as some of the results presented here are also
relevant to those other models.

One of the purposes of this paper is to give a careful definition of
``good space--time'' under which Theorem \ref{T0} holds, let us
therefore start with the complete basics. A couple $(M,g)$ will be
called a space--time if $M$ is a smooth, connnected, Hausdorff,
paracompact manifold of dimension $4$ and $g$ is a smooth
non--degenerate tensor field of Lorentzian signature, say $-+++$. We
shall also assume that there exists a smooth Killing vector field $X$
on $M$. To the conditions listed here we shall need to add several
more conditions, each of which will be discussed below in a separate
section:

\subsection{Non--degenerate horizons vs. bifurcation surfaces}
The uniqueness Theorem of Ruback \cite{Ruback} ({\em cf.\/} also Simon
\cite{Simon2}[Section 4] or Masood--ul--Alam \cite{MuA} for some
simplifications of the argument, and Carter \cite{CarterHI} and
references therein for previous results on this problem) is often
referred to as a uniqueness Theorem for static non--degenerate
electro--vacuum black holes. This description is incorrect and
misleading. Recall that given a space--time $(M,g)$ with a Killing
vector field $X$ one defines the {\em Killing horizon} $\cNX$ as the
set of points on which $X$ is {\em null} and {\em non--vanishing}.
[In this definition we do {\em not} assume that $X$ is necessarily
timelike at inifinity]. By an abuse of terminology, a connected
component of $\cNX$ will sometimes also be called a Killing horizon.
It is well known and in any case easily seen that there exists on
$\cNX$ a function $\kappa$, called {\em surface gravity}, which is
defined by the equation
\begin{equation}
\label{definekappa}
{1\over 2} \nabla ^a(X^bX_b)\Big|_\cNX=-\kappa X^a\ .
\end{equation}
$\kappa$ is known to be constant over every connected component of
$\cNX$ for electro--vacuum space--times \cite{BCH} ({\em cf.\/} also
\cite{KayWald} for a simple proof in the bifurcate-horizons case). A
connected component $\cNX_a$ of $\cNX$ is said to be {\em degenerate} if
$\kappa|_{\cNX_a}\equiv 0$. Let us mention that for Kerr--Newman
metrics we have $\kappa\ne 0$ as long as $M^2>Q^2+L^2$, where $M$ is
the ADM mass of the metric, $L$ its ADM angular momentum and $Q$ the
total charge of the electric field. On the other hand the
Majumdar--Papapetrou black holes of ref.\ \cite{HartleHawking} ({\em
  cf.\/} also Appendix \ref{Veron}) have
$\kappa\equiv0$ throughout the Killing horizon.

Consider the Schwarzschild--Kruskal--Szekeres space--time $(M,g)$, let
$X$ be the standard Killing vector field which equals
$\partial/\partial t$ in the asymptotic regions. Recall that the
Killing horizon $\cNX$ of 
$X$ has four connected components, such that the set
$\cSX\equiv\overline \cNX\setminus\cNX$, where $\overline \cNX$
denotes the topological closure of $\cNX$, is a smooth
two--dimensional embedded submanifold of $M$. The Killing vector $X$
vanishes on $\cSX$.  By definition, such a surface will be called a
{\em bifurcation surface of a bifurcate Killing horizon} ({\em cf.\/}
\cite{Boyer} for a justification of this terminology). Thus, given a
non--identically vanishing Killing vector field $X$, a bifurcation
surface is a smooth two--dimensional compact embedded surface on which
$X$ vanishes. If such a surface $\cSX$ exists, then every connected
component of $\cNX$ such that $\overline \cNX\cap \cSX\ne\emptyset$ is
{\em necessarily} a non--degenerate Killing horizon \cite{Boyer} ({\em
  cf.\/} also \cite{KayWald}).  It follows that the existence of a
bifurcation surface $\cSX$ implies that of a non--degenerate Killing
horizon, {\em but the converse is not true in general}. A rather
trivial example is obtained by removing $\cSX$ from a space--time
which contains such a surface\footnote{ Note that this example shows
  that Theorem 5.1 of (the otherwise excellent and in many respects
  fundamental) Ref.\ \cite{CarterLesHouches} is wrong.}.  A somewhat
less trivial example is that of any vacuum space--time with a smooth
compact non--degenerate Cauchy horizon with a Killing vector say,
spacelike on a Cauchy surface ({\em e.g.}, the Misner model for
Taub--NUT space--times, or the Taub--NUT space--times themselves
\cite{HE}).

It is natural to look for conditions under which the existence of a
non--degenerate Killing horizon does indeed imply that of a
bifurcation surface. Some results in this direction have recently been
obtained by R\'acz and Wald \cite{RW}, who have shown that there seems
to be no local obstruction to the existence of a bifurcation surface,
when a non--degenerate Killing horizon is present. More precisely,
assuming that the non--degenerate Killing horizon has a global
cross--section ({\em i.e.}, a two dimensional submanifold which is
intersected by every generator of the horizon precisely once), R\'acz
and Wald show\footnote{In our discussion of the results of \cite{RW}
we assume that the electro--vacuum equations are satisfied. This
hypothesis is not made in \cite{RW}, which allows for non--constant
$\kappa$. It is shown in \cite{RW} that such a possibility leads to
the existence of a ``parallel propagated singularity" of the curvature
tensor.} that whenever a bifurcation surface does not exist, then one
can make a {\em local} extension which contains one.  Recall now that
the difference between a local extension and a ``real one'' is the
following: to obtain an extension of a space--time $(M,g)$ one
constructs a space--time $(\hat M,\hat g)$ and an embedding
$i:M\to\hat M$ such that $i^*\hat g = g$ and $i(M)\ne \hat M$; for
{\em local} extensions one considers a subset $\cU\subset M$, and one
constructs\footnote{In the definition of local extension one sometimes
adds some supplementary conditions on $\cU$ and $\hat\cU$ which are
irrelevant for the discussion here, {\em cf.\ e.g.\/}
\cite{HE,Racz,Beem,Clarke} and references therein.} an extension
$(\hat\cU,\hat g)$ of $(\cU,g|_\cU)$.  The problem here is that
sometimes there is no way of patching $(M,g)$ with $(\hat\cU,\hat g)$
to obtain either a manifold ({\em i.e.}, Hausdorffness of the
resulting topological space might be violated) or a continuous metric.
[For example, extensions where continuity of the metric and
Hausdorffness cannot be simultaneously ensured can be constructed in
the vacuum Einstein class using the polarized Gowdy metrics,
exploiting the asymptotic behaviour of the metric near the $t=0$ set
described in \cite{ChIM}. Examples of local extensions in the 
Killing--horizon--context which cannot be turned into ``real ones''
have been constructed by Wald\footnote{R.\ Wald, private
communication.} (without, however, satisfying any field equations or
energy inequalities).]

We wish to emphasize, that {\em the uniqueness theorems of
\cite{Ruback,Simon2,MuA} implicitly assume\footnote{It is far from
being obvious that in a general static space--time $(M,g) $ there will
exist a hypersurface satisfying the conditions of
\cite{Ruback,Simon2,MuA}. In electro--vacuum it is not too difficult
to show that the conditions of \cite{Ruback,Simon2,MuA} are equivalent
to the existence of a bifurcation surface in $M$ (assuming maximal
global hyperbolicity, asymptotic flatness, etc.). [By {\em maximal
globally hyperbolic\/} we always mean maximal in the class of globally
hyperbolic space--times.]} the existence of a compact bifurcation
surface} in the space--time under consideration. It must therefore be
stressed that the existing uniqueness theory is that of stationary
electro--vacuum space--times with bifurcation surfaces and {\em not}
that of stationary electro--vacuum space--times with a non--degenerate
horizon.  A way of obtaining uniqueness results in the latter class of
space--times would be to prove the following, or some variation
thereof\footnote{This point has already been stressed by R\'acz and
Wald in \cite{RW}, and was indeed the main motivation for the work in
\cite{RW}.}:
\begin{Conjecture}
\label{C1}
Let $(M,g)$ be a maximal globally hyperbolic, asymptotically flat,
electro--vacuum space--time with Killing vector field $X$ which is
timelike in the asymptotically flat regions. Suppose that $(M,g)$ is
{\em not} the Minkowski space--time, let $Z$ be a Killing vector field
on $M$ (perhaps, but not necessarily coinciding with $X$), and suppose that $M$
contains a non--empty Killing horizon $\cN{[Z]}$.  Then for every
non--degenerate connected component $\cN_a$ of $\cN[Z]$, the set
$S_a\equiv \overline{\cN_a}\setminus \cN_a$ (where $\overline{\cN_a}$
is the topological closure of ${\cN_a}$) is a non--empty, compact,
embedded, smooth submanifold of $M$.
\end{Conjecture}

Let us emphasize that {\em no uniqueness results have been established
  so far for space--times with a degenerate Killing horizon
  ($\kappa=0$).} [Thus, as of today the definition of ``good'' in
Theorem \ref{T0} includes the notion of non--degeneracy. With that
condition the conclusion of that Theorem can clearly be strenghtened
to exclude the Majumdar--Papapetrou black--holes, {\em cf.\/} Theorem
\ref{T3} below.] It is customary to rule out the degenerate horizons
as physically uninteresting, as their defining property is of unstable
character.  [Moreover, they can perhaps be discarded as physically
irrelevant by thermodynamical considerations, as $\kappa$ is related
to some kind of ``temperature of the black hole'' \cite{BCH} ({\em
  cf.\/} also \cite{KayWald} and references therein).] For the sake of
mathematical completeness one would nevertheless like to have a
classification of the degenerate cases as well:

\begin{Problem}
\label{P1}
Classify all the maximal, electro--vacuum
space--times $(M.g)$ satisfying the following:
\begin{enumerate}
\item in $M$ there exists a Killing vector field $X$ which is timelike
  in the asymptotic regions;
\item $M$ contains  a  partial Cauchy surface $\Sigma$ with
  asymptotically flat ends which is a complete Riemannian manifold
  with respect to the induced metric;
\item there are no naked singularities in $J^+(\Sigma)$; and finally
  \item $M$ contains degenerate Killing horizons.
\end{enumerate}
\end{Problem}

It seems that the only known space--times satisfying the above are
those Majumdar--Papapetrou space--times
\cite{Majumdar,Papapetrou,HartleHawking} which contain a finite number
of black holes; the V\'eron solutions with an infinite number of black
holes described in Appendix \ref{Veron} are probably excluded by the
condition of 
absence of naked singularities, {\em cf.\/} Appendix \ref{Veron}.  One
of the difficulties which might arise here is, that there is no reason
for a degenerate Killing horizon to be smooth (note that
\eq{definekappa} guarantees the smoothness of the Killing horizon when
$\kappa\ne 0$ and when the space--time metric is smooth).

It seems, moreover, that no globally hyperbolic asymptotically flat
electro--vacuum space--times are known which possess a {\em complete
Cauchy surface} and a {\em degenerate} Killing horizon: In the
Majumdar--Papapetrou black--holes analyzed in \cite{HartleHawking} the
Killing horizon is also a Cauchy horizon for the (complete) ``static'
partial Cauchy surfaces $t=\mbox{const}$. [Note, moreover, that those
``static" partial Cauchy surfaces are not asymptotically flat in the
usual sense: in addition to the asymptotically flat ends they contain
"infinite asymptotically cylindrical necks".]  On the other hand,
those partial Cauchy surfaces which intersect the Killing horizon
cannot probably be complete because of the singularities present.

To close this Section it should be admitted that there is not much
evidence that the Majumdar--Papapetrou space--times play the
role advertised in Theorem \ref{T0}. The author bases his belief on
the analysis of \cite{GibbonsHull,GHHP} ({\em cf.\/} also \cite{Tod}),
where an argument is given that (under some yet--to--be--specified
conditions) for any electro--vacuum space--time we must have $M\ge
\sqrt{Q^2+P^2}$, where $M$ is the ADM mass and $Q,P$ are the global
electric and magnetic charges, with the bound being saturated
precisely by the Majumdar--Papapetrou space--times.  The reader
should, however, note that the local analysis of \cite{GibbonsHull,Tod}
should be complemented by a global one, related to the questions
raised above of existence of appropriately regular space--like
surfaces, etc. To the author's knowledge this has not been done yet.

\subsection{Killing vectors vs.\ isometries}
\label{isometries}
In general relativity there exist at least two ways for a solution to
be symmetric: there might exist
\begin{enumerate}
  \item\label{field} a Killing vector field $X$ on the space--time
$(M,g)$, or there might exist \item\label{globalaction} an action of a
(non--trivial) connected Lie group $G$ on $M$ by isometries.
\end{enumerate}

Clearly \ref{globalaction} implies \ref{field}, {\em but \ref{field}
  does not need to imply \ref{globalaction}} (remove {\em e.g.\/}
points from a space--time on which an action of $G$ exists).  In the
uniqueness theory, as presented {\em e.g.\/} in
\cite{HE,CarterCargese}, one {\em always} assumes that an action of a
group $G$ on $M$ exists. This is equivalent to the statement, that the
orbits of all the (relevant) Killing vector fields are complete. When
trying to classify space--times with
      Killing vector fields, as in Theorem \ref{T0}, one
      immediately faces the question whether or not the orbits thereof
      are complete. It is worthwile emphasizing that there is a
      constructive method of producing space--times with Killing
      vectors, by solving a Cauchy problem:

\begin{Theorem}
\label{TKilling}
Let $(\Sigma,\gamma,K,A,E)$ be initial data\footnote{Here $\Sigma$ is
a three--dimensional manifold, $\gamma$ is a Riemannian metric on
$\Sigma$, $K$ is a symmetric two--tensor on $\Sigma$, $A$ and $E$ are
vector fields on $\Sigma$. Moreover all the fields are assumed to
satisfy the electro--vacuum constraint equations.} for electro--vacuum
Einstein equations, let $(M,g)$ be any globally hyperbolic
(electro--vacuum) development thereof.  Suppose that there exists a
vector field $\hat X$ defined\footnote{The vector field $\hat X$ has
been introduced here purely for notational convenience.  Alternatively
one can rewrite eqs.\ \eq{eq1}--\eq{eq3} purely in terms of initial
data for eq.\ \eq{eqstar} below.} in a neighbourhood of $\Sigma$ such
that the following equations hold on $\Sigma$:
\begin{eqnarray}
\label{eq1} &{
  \Big\{\nabla_\mu \hat X_\nu+\nabla_\nu \hat
  X_\mu\Big\}\Big|_{\Sigma}=0, \qquad
  \Big\{\nabla_\alpha\Big(\nabla_\mu \hat X_\nu+\nabla_\nu \hat
  X_\mu\Big)\Big\}\Big|_{\Sigma}=0\ , }& \\
\label{eq3} & { {\cal L}_{\hat X}
  F_{\mu\nu}\Big|_{\Sigma}=0 
 \ .} &
\end{eqnarray}
Here $\nabla $ is the covariant derivative of $g$, ${\cal L}$ denotes
a Lie derivative and $F_{\mu\nu}$ is
the electromagnetic field tensor.
Then there exists on $M$ a Killing vector field $X$ which coincides
with $\hat X$ on $\Sigma$.
\end{Theorem}

{\sc Remarks:} \begin{enumerate}
\item Let us note that eqs.\ \eq{eq1}--\eq{eq3} are necessary, as they
  automatically hold if $\hat X$ is a Killing vector which moreover
leaves the electromagnetic field invariant.  \item It must be
emphasized that equations \eq{eq1}--\eq{eq3} need to hold {\em on
$\Sigma$ only}. [These equations can be thought of as constraint
equations for the initial data for a Killing vector
field.]  In other words, it is sufficient to satisfy the Killing
equations on $\Sigma$ to obtain a solution of the Killing equations on
{\em any} (not necessarily maximal) globally hyperbolic development
thereof.
\item The vacuum equivalent of Theorem \ref{TKilling} is well known
  \cite{MoncriefKilling,FMM,SCC}.
\item It would be of interest to obtain an equivalent of Theorem
  \ref{TKilling} for Einstein--Yang--Mils equations ({\em cf.\/}
  \cite{Arms} for some related results).
\end{enumerate}

{\sc Proof:} Let $X^\mu$ be defined as the unique solution of the
problem \bea & \Box X^\mu = -{R^\mu}_\nu X^\nu\ , & \label{eqstar}\\ &
X^\mu\Big|_\Sigma = \hat X^\mu, \qquad \nabla_\alpha
X^\mu\Big|_\Sigma = \nabla_\alpha \hat X^\mu\ ,  \nonumber \eea define
$$
A_{\mu\nu} = \nabla_\mu X_\nu+\nabla_\nu X_\mu\ .
$$ From \eq{eqstar} and from the Einstein--Maxwell equations one
derives the following system of equations \bea\label{eqstar1} & \Box
A_{\mu\nu} = -2{\cal L}_X R_{\mu\nu} + 2
{R^\lambda}_{(\mu}A_{\nu)\lambda} + 2{{ R_\mu}^{\alpha\beta}}_\nu
A_{\alpha\beta}\ , & \\\label{eqstar2} & \nabla^\mu{\cal
  L}_XF_{\mu\nu} = \nabla_\alpha F_{\beta\nu}A^{\alpha\beta} +
F_{\alpha\beta}\nabla^\alpha{A^\beta}_\nu \ .& \eea Note that because
of the Einstein equations the tensor field ${\cal L}_X R_{\mu\nu}$ can
be expressed as a linear combination of $A_{\alpha\beta}$ and ${\cal
  L}_XF_{\alpha\beta}$. The initial data for \eq{eqstar1}--\eq{eqstar2}
vanish by \eq{eq1}--\eq{eq3}, and the vanishing of $A_{\alpha\beta}$
and of ${\cal L}_XF_{\mu\nu}$ follows.  \hfill$\Box$

Recall now that given a Cauchy data set $(\Sigma,\gamma,K,A,E)$ for
electro--vacuum Einstein equations, there exists a {\em unique up to
  isometry} vacuum space--time $(M,g)$, which is called the {\em
  maximal globally hyperbolic vacuum development of
  $(\Sigma,\gamma,K)$}, with an embedding $i:\Sigma\rightarrow M$ such
that $i^*g=\gamma$, and such that $K$ corresponds to the extrinsic
curvature of $i(\Sigma)$ in $M$ \cite{ChG}. $(M,g)$ is {\em
  inextendible} in the class of globally hyperbolic space--times with
a vacuum metric. This class of space--times is highly satisfactory to
work with, as they can be characterized by their Cauchy data induced
on some Cauchy surface. Moreover, the property of maximality seems to
be a natural notion of completeness for globally hyperbolic
space--times, and it is of interest to enquire about completeness of
Killing orbits in such space--times.  Before discussing that question,
it seems appropriate to introduce some definitions:

\begin{Definition}
\label{D1}
We shall say that an initial data set $(\veps,\gamma,K,A,E)$ for
electro--vacuum Einstein equations is asymptotically flat if
$(\veps,\gamma)$ is a complete connected Riemannian manifold (without
boundary), with $\Sigma$ of the form
\begin{equation}
\label{topological}
\veps=\veps_{\rmint} \bigcup^I_{i=1}\veps_i, \end{equation} for some
$I<\infty$.  Here we assume that $\veps_{\rmint}$ is compact, and each
of the ends $\veps_i$ is diffeomorphic to $\R^3\setminus B(R_i)$ for
some $R_i>0$, with $B(R_i)$ --- coordinate ball of radius $R_i$. In
each of the ends $\veps_i$ the fields $(g,K,A,E)$ are assumed to
satisfy the following inequalities (after performing a duality
rotation of the electromagnetic field, if necessary)
\begin{equation}
\label{falloff}
|g_{ij}-\delta_{i}^j|+|r\partial_k g_{ij}|+|rK_{ij}| +|A_i| +
|r\partial_i A_j| +|rE_i| \le C r^{-\alpha}\ ,
\end{equation}
for some positive constant $C$ and some $\alpha >0$, with
$r=\sqrt{\sum (x^i)^2}$.
\end{Definition}

To motivate the next definition, consider a space--time with some
number of asymptotically flat ends, and with a black hole region. In
such a case there might exist a Killing vector field defined in, say,
the domain of outer communication ({\em cf.\/} the next Section for
a definition) of the asymptotically flat ends. It could, however,
occur, that there is no Killing vector field defined on the whole
space--time --- an example of such a space--time has been considered
by Brill \cite{Brill}, in his construction of a space--time in which
no asymptotically flat maximal surfaces exist. Alternatively, there
might be a Killing vector field defined everywhere, however, there
might be some non-asymptotically flat ends in $\Sigma$. [As an
example, consider a spacelike surface in the
Schwarzschild--Kruskal--Szekeres space--time in which one end is
asymptotically flat, and the second is ``asymptotically
hyperboloidal''.] In such cases one would still like to claim that the
orbits of $X$ are complete in the exterior region.  To accomodate such
behavior we introduce the following:

\begin{Definition}
\label{D2}
Consider a stably causal Lorentzian manifold $(M,g)$ with an achronal
spacelike surface $\hat\Sigma$.  Let $\Sigma \subset \hat\Sigma$ be a
connected submanifold of $\hat \Sigma$ with smooth compact boundary
$\partial\Sigma$, and let $(\gamma,K)$ be the Cauchy data induced by
$g$ on $\Sigma$.  Suppose finally that there exists a Killing vector
field $X$ defined on $\D(\Sigma)$ (here $\D(\Sigma)$ denotes the
domain of dependence of an achronal set $\Sigma$; we use the
convention in which $\D(\Sigma)$ is an {\em open} set). We shall say that
$(\Sigma,\gamma,K,A,E)$ are Cauchy data for an asymptotically flat
exterior region in a (non--degenerate) black--hole space--time
if the following hold:
\begin{enumerate}
\item The closure $\bar \Sigma\equiv \Sigma\cup\partial\Sigma$ of
  $\Sigma$ is of the form (\ref{topological}), with $ \veps_{\rmint}$
  and $\veps_i$ satisfying the topological requirements of Definition
  \ref{D1}.,
\item $(\Sigma,\gamma,K,A,E)$ satisfy the fall--off requirements of
  Definition \ref{D1}.

\item \, [From the Killing equations it follows that $X$ can be
  extended by continuity to $\overline{ \D(\Sigma)}$.] We shall
  moreover require that $X$ be tangent to $\partial \Sigma$.
\end{enumerate}
\end{Definition}

The above definition allows for space--times in which 
$\Sigma$ is a surface with boundary, the boundary in question being a
bifurcation surface of a Killing horizon.  The notion of {\em
  non--degeneracy} referred to in definition \ref{D2} above is related
to the non--vanishing of the surface gravity of the
horizon:  Indeed, it follows from \cite{RW} that 
in situations of interest the behaviour described in Definition
\ref{D2} can only occur if the
surface gravity of the horizon is constant  on the horizon, and does
not vanish.

The following is a straightforward generalization\footnote{The
  remaining results of \cite{Chorbits} can be similarly generalized to
  the electro--vacuum  case.} of the Theorem proved in \cite{Chorbits},
no details will be given:

\begin{Theorem}
\label{Torbits}
Let $(M,g)$ be a smooth, electro--vacuum, maximal globally hyperbolic
space--time with an achronal spacelike hypersurface $\Sigma$ and with
a Killing vector $X$ (defined perhaps only on $\D(\Sigma)$) such that
$X$ approaches\footnote{Theorem \ref{Torbits} holds true for any
  Killing vector field, provided that asymptotic conditions somewhat
  stronger than those of Definition \ref{D1} are assumed, {\em cf.\/}
  \cite{Chorbits} for details.} a non--zero multiple of the unit
normal to $\Sigma$ as $r\to\infty$.  Suppose that either
\begin{enumerate}
\item the data set $(\Sigma,\gamma,K,A,E)$ is asymptotically flat, or
\item $(\Sigma,\gamma,K,A,E)$ are Cauchy data for an asymptotically
  flat exterior region in a (non--degenerate) black--hole space--time.
\end{enumerate}
Then the orbits of $X$ are complete in $\D(\Sigma)$.
 \end{Theorem}

Consider then a stationary black hole space--time $(M,g)$ in which an
asymptotically flat Cauchy surface exists but in which the Killing
orbits are {\em not} complete: Theorem \ref{Torbits} shows that
$(M,g)$ can be enlarged to obtain a space--time with complete Killing
orbits.

In conclusion, the results  presented in this Section
show that the hypothesis  of completeness of Killing orbits  usually
made in uniqueness Theorems is unnecessary, as long as one restricts
oneself to maximal globally hyperbolic space--times with well behaved
Cauchy surfaces.

\subsection{Asymptotic flatness, stationarity}
\label{asymptotic}
There are at least three different ways of defining asymptotic
flatness:
\begin{enumerate}
\item via existence of an asymptotically flat Cauchy surface, or
\item via existence of asymptotically Minkowskian coordinates, and
  finally
\item using conformal techniques.
\end{enumerate}
More precisely, let $(M,g)$ be an electro--vacuum space--time. We
shall say that a submanifold\footnote{We use the (PDE motivated)
  convention that a submanifold $\Sigma$ with boundary does {\em not}
  include its boundary $\partial \Sigma\equiv\overline\Sigma\setminus
  \Sigma$.} $\hat \Sigma$ with boundary is {\em an asymptotically flat
  three--end in $M$} if $\hat \Sigma$ is diffeomorphic to $\R^3\setminus
B(R)$ for some $R>0$, where $B(R)$ denotes a closed coordinate ball of
radius $R$, and in the local coordinates on $\hat \Sigma$ the fields
$(g,K,A,E)$ satisfy the fall--off conditions \eq{falloff} of
Definition \ref{D1}.

We shall say that an open submanifold $\hat M\subset M$ is {\em an
  asymptotically flat stationary four--end of $M$} if $\hat M$ is
diffeomorphic to $\R\times (\R^3\setminus B(R))$, and in the local
coordinates on $\hat M$ the metric $g_{\mu\nu}$, the electromagnetic
potential $A_\mu$ and the elecromagnetic field $F_{\mu\nu}$
satisfy (after perfoming a duality rotation of the electromagnetic
field, if necessary)
\begin{eqnarray}
\label{spacetimefalloff}
& |g_{\mu\nu}-\eta_{\mu\nu}|+|r\prt _i g_{\mu\nu}| +|A_\mu|
+|rF_{\mu\nu}|\le Cr^{-\alpha}\ ,&
\\
\label{spacetimefalloff2}
& \prt _t g_{\mu\nu}= \prt _tF_{\mu\nu}=0\ ,&
\end{eqnarray}
for some constants $C,\alpha>0$. Here we have $r=\sqrt{\sum (x^i)^2}$,
as before.

Clearly a space--time with an asymptotically flat stationary four--end
$\hat M$ also contains an asymptotically flat three--end $\hat \Sigma$
and a Killing vector which is timelike on $\hat \Sigma$, {\em but the
  converse needs not to be true.} This is due to the fact that a
timelike Killing vector field $X$ defined on $\hat \Sigma$ might
asymptotically approach a null (rather than timelike) Killing vector
as $r$ goes to infinity, say $X\to_{r\to\infty}\prt _t-\prt _z$. An explicit
example of such a space--time (not satisfying any reasonable field
equations or energy conditions) can be found in the Appendix A of
\cite{ChWald}. When imposing electro--vacuum field equations such a
behaviour seems to be rather improbable.
For the sake of completeness
of understanding space--time with Killing vectors which are timelike
in the asymptotic regions it would be of interest to prove the
following:
\begin{Conjecture}
\label{nonullconjecture}
Let $(M,g)$ be an electro--vacuum space--time with an asymptotically
flat three--end $\hat\Sigma$ and a Klling vector field $X$ which is
timelike on $\hat\Sigma$. After performing a boost of $\hat \Sigma$ if
necessary, $X$ approaches a non--zero multiple of the unit normal to
$\hat\Sigma$ as $r$ goes to infinity.
\end{Conjecture}

The following gives a plausibility argument for Conjecture
\ref{nonullconjecture}: Suppose that the Killing vector $X$
asymptotically approaches a null vector at $i^o$. Under these
circumstances one would expect the ADM four--momentum to be parallel
to the Killing vector,\footnote{This result should probably follow by,
  {\em e.g.}, a repetition of the analysis of \cite{AAM}.} hence null.
This is, however, not possible when energy conditions are satisfied
\cite{AshtekarHorowitz}. We find it likely that a proof of Conjecture
\ref{nonullconjecture} can be given by filling in the details in this
argument.

Under the conditions and conclusions of Conjecture
\ref{nonullconjecture} it is rather easy to show that $M$ will also
contain an asymptotically flat stationary four--end $\hat M$, provided
that the orbits of $X$ through $\hat \Sigma$ are complete ({\em cf.\
  e.g.\/} \cite{ChWald}[Appendix A]). In this case $\hat M$ can be
defined by the equation
\begin{equation}
\label{fourend}
\hat M=\bigcup_{t\in\R}\phi_t(\hat\Sigma)\ ,
\end{equation}
where $\phi_t$ is the flow generated by the Killing vector field $X$.
This together with the discussion of the previous Section shows the
equivalence of the "$3+1$ Definition" and the "4--dim Definition" of
asymptotic flatness for maximal globally hyperbolic electro--vacuum
space--times with an asymptotically timelike Killing vector $X$,
modulo the proviso of the validity of the conclusion of Conjecture
\ref{nonullconjecture}.

As far as the conformal approach is concerned, we have the following:

\begin{Proposition}
\label{PAF1}
Suppose that an electro--vacuum space--time $(M,g)$ contains an
asymptotically flat stationary four--end $\hat M$. Then $M$ admits a
conformal completion satisfying the completeness requirements of
\cite{GHor}.
\end{Proposition}

\proof A bootstrap of the stationary field equations in $\hat M$ shows
that one can find a coordinate system and an electromagnetic gauge in
which \eq{spacetimefalloff2} holds and moreover the fields
satisfy\footnote{In vacuum this observation has been independently
  done by D.\ Kennefick and N.\ O'Murchadha \cite{KOM}.}
\begin{equation}
\label{spacetimefalloff3}
|g_{\mu\nu}-\eta_{\mu\nu}|+r|\prt _i g_{\mu\nu}| +|A_\mu|
+|rF_{\mu\nu}|\le Cr^{-1}\ , \end{equation} The results of Ref.\
\cite{Simon1} and \eq{spacetimefalloff3} show that $\hat M$ admits a
smooth conformal completion at $i^o$. The Appendix to \cite{DS} gives
then an explicit construction of the conformal completion at null
infinity.  \hfill $\Box$

A converse of Proposition \ref{PAF1} can be proved again under some
provisos, including the validity of an appropriately modified version
of Conjecture \ref{nonullconjecture}.
Indeed, if the Ricci tensor falls off fast enough (in the sense of the
note added in proof (3) of \cite{AX}) in the asymptotic end in
question near a connected component$\hat{\scri}$ of $\scri$ (and this
decay probably follows from the peeling property of the
electromagnetic field) 
then Bondi coordinates
near $\hat{\scri}$, and subsequently asymptotically Minkowskian
coordinates near $\hat{\scri}$ can be constructed. If the Killing
vector does not approach an asymptotically null vector,
then this construction gives an asymptotically flat stationary
four--end $\hat M$.

We wish to stress that the field equations played a significant role
in the discussion above. Recall that one does {\em not} expect a
general asymptotically Minkowskian space--time to admit smooth
conformal completions \cite{Fock,ChKl,Madore,PRL,ChrMCS}. As shown
in Appendix \ref{obstruction}, the same is true for general
asymptotically Minkowskian stationary space--times when one does not
impose any field equations.  When a stationary space--time admits a
$\Scri$ which is merely polyhomogeneous rather than smooth, {\em
  i.e.}, when the metric has $r^{-j } \log^i r$ terms in its
asymptotic expansion for large $r$, various technical difficulties
arise when asymptotic flatness is defined in terms of a conformal
completion and several of the results discussed in {\em e.g.\/}
\cite{HE} require reexamination.  It follows that the question of
equivalence of the conformal definition of asymptotic flatness with
the other ones requires a case by case analysis for each matter model.
All these difficulties are, however, avoided, when using the
definitions of asymptotic flatness based on existence of appropriate
coordinate systems, as discussed above.

It should be noted that the question of definition of asymptotic
flatness is related to that of the definition of the black--hole region. In
\cite{HE,CarterLesHouches} one considers connected components
$\Scri_i^\pm$ of $\Scri$ and one defines the {\em black hole region}
$\cB_i $ as $M\setminus J^-({\scri^+_i})$. Similarly the {\em white
  hole region} $\cW_i $ is defined as $M\setminus J^+({\scri^-_i})$,
and the {\em domain of outer communication} $\langle\langle \scri_i
\rangle\rangle$ is defined as $J^-({\scri^+_i})\cap J^+({\scri^-_i})$.
On the other hand, in \cite{ChWald} one considers an asymptotically
flat stationary four--end $M_i$ and then the black hole, white hole,
etc., are defined as
\begin{eqnarray}
&\label{blackhole}\cB_i\equiv M\setminus J^-({M_i}), \qquad
\cW_i\equiv M\setminus J^+({M_i})\ ,&
\\ 
\label{doc}
 &
\langle\langle \scri_\Sigma \rangle\rangle\equiv \Big\{\cup_i
J^-({M_i})\Big\}\bigcap\Big\{\cup_i J^+({M_i}) \Big\}
\ .
&
\end{eqnarray}
Here the $M_i$'s are defined as in \eq{fourend}, starting from the
asymptotic three--ends $\Sigma_i$  of $\Sigma$.
 As discussed in
\cite{ChWald}, these definitions coincide with the ones based on
conformal completions in vacuum; from what it said here it follows
that this is also true in the electro--vacuum case modulo some
provisos discussed above.
The advantage of the conformal definition of black hole, etc., is that
it carries over immediately to the non--stationary cases\footnote{Note
  that the conformal definitions of black hole, etc., still make  sense
  with conformal completions of poor differentiability. For such
  completions, however, various properties of $\cB$, etc., should be
  carefully reexamined.}.

\section{Two uniqueness Theorems}

\subsection{The angular velocity of bifurcation surfaces of
  non--degenerate Killing horizons}
\label{Wald}
Before presenting a uniqueness theorem for black holes, let us report
here the following unpublished result of Wald which allows us to
define the angular velocity for bifurcation surfaces of
non--degenerate Killing horizons, and (in the case of non--vanishing
angular velocities) a preferred (non--trivial) Killing vector $Y$ with
{\em periodic} orbits (no field equations are assumed below):

\begin{Proposition}
\label{I3.1}
Consider a space--time $(M,g)$ with a Killing vector field $X$ with
complete orbits which contains an asymptotically flat stationary
four--end $\hat M$. Suppose moreover that there exist in $M$ a
compact, smooth, two dimensional (not necessarily connected)
submanifold $S$ with the following properties: for every connected
component $S_j$, $j=1,\ldots, J$ of $S$ there exists a Killing vector
$Z_j$ with complete orbits in $M$ which vanishes on $S_j$.  Then
either $X$ coincides with all the $Z_j$'s, in which case we set
$\Omega_j=0$ for all $j$, or there exists on $M$ a Killing field $Y$
such that

1. $Y$ commutes with $X$,

2. $Y$ is complete and has periodic orbits with period $2\pi$, and

3. for each $j = 1, \ldots, J$ there exists $\Omega_j \in \R$ such
that, rescaling $Z_j^a$ if necessary, we have $Z_j^a = X^a + \Omega_j
Y^a$.
\end{Proposition}
It might be of some interest to note, that the arguments of Wald show
moreover the following:

1. If any connected component of $S$, say $S_1$, is not diffeomorphic
to a torus or a sphere, then every Killing vector has to vanish on
$S_1$ (this has already been observed in \cite{GH}). It follows that
in such a case $(M,g)$ can have (up to proportionality) at most one
Killing vector the orbits of which are complete.

2. If any connected component of $S$, say $S_1$, is diffeomorphic to a
sphere, then either there are at most two linearly independent Killing
vectors with complete orbits in $M$, or $M'$ is static, spherically
symmetric, and the asymptotically stationary Killing vector $X$
vanishes on $S$.

3. If a connected component of $S$, say $S_1$, is diffeomorphic to a
torus, then any Killing vector with complete orbits must have periodic
orbits on $S_1$; moreover $(M,g_{ab})$ can have at most two linearly
independent Killing vectors with complete orbits.

Proposition \ref{I3.1} can be used as a starting point for a
classification of stationary space--times with bifurcation surfaces.

The constant $\Omega_i$ defined in Proposition \ref{I3.1} will be
called {\em the angular velocity} of the $i$'th connected component of
the black hole.

\subsection{A uniqueness Theorem for black holes}
\label{uniqueness}
In this Section we shall present a version of Theorem \ref{T0}. The
main steps of the proof are the Sudarsky--Wald staticity theorem
\cite{SW2} ({\em cf.\/} also \cite{SW1}) and the Bunting --
Masood--ul--Alam -- Ruback \cite{BMuA,Ruback} uniqueness theorem for
static electro--vacuum black holes. Let us start with a Definition:

\begin{Definition}[Condition ${\cal C}1$]
  \label{DC1} A quadruple $(M,g,X,\Sigma)$ will be said to {\em
    satisfy the condition $\cC 1$} if $(M,g)$ is a maximal globally
  hyperbolic electro--vacuum space--time with electromagnetic field
  $F$ and if moreover the following conditions are satisfied:

  1. $\Sigma$ is a simply connected\footnote{\label{simply}The
    hypothesis of simple connectedness of $\Sigma$ is used in the
    Theorems below to ensure the existence of a global gauge in which
    \eq{new.1} holds.  This hypothesis is therefore unnecessary in
    vacuum, or in situations in which one knows {\em a priori} ({\em
      e.g.}, by assumption, as in \cite{SW1,SW2}) that a global gauge
    satisfying \eq{new.1} exists.} spacelike hypersurface in $M$
  satisfying the requirements of Definition \ref{D2}.

  2. $X$ is a Killing vector field defined on $\cD(\Sigma)$ such that
${\cal L}_X F=0$.  Moreover there exist constants $\alpha_i\in\R $
such that on every asymptotic three--end $\Sigma_i$ of $\Sigma$ we
have (after performing a Lorentz ``boost" of $\Sigma_i$ if necessary)
\begin{equation} X\Big|_{\Sigma_i}\to_{r\to\infty}\alpha_i n\ ,
\label{C1.1} \end{equation} where $n$ is the unit future directed
normal to $\Sigma$. We shall normalize\footnote{\label{samesign}The
non--vanishing of the $\alpha_i$'s for a non--trivial Killing vector
field $X$ is a well known consequence of the Kiling equations.} the
$\alpha_i$'s so that $ \alpha_1 = 1 $.

      3. For every connected component $\prt\Sigma_a$ of $\prt \Sigma$
      there exists a Killing vector $Z_a$ defined on
      $\overline{\cD(\Sigma)}$ which vanishes on $\prt\Sigma_a$. We
      also require ${\cal L}_{Z_a}F=0$.

      4. Let the domain of outer communication $ \langle\langle \scri_\Sigma
      \rangle\rangle$ be defined by \eq{doc}. We shall require that
      \begin{equation} \cD(\Sigma)\subset \langle\langle \scri_\Sigma
        \rangle\rangle\ .
\label{C1.3}
\end{equation}
In other words, $\Sigma$ and its domain of dependence $\cDs$ lie
entirely outside the black hole and the white hole regions.
    \end{Definition}


If $(M,g,X,\Sigma)$ satisfy the condition $\cC 1$, then every Killing
vector defined on $\cDs$ has complete orbits. We can consequently
use\footnote{\label{fnote2}Strictly speaking, Proposition \ref{I3.1}
  has been formulated in a way which assumes the existence of Killing
  vectors defined globally on $M$. It can, however, be seen that its
  assertions hold true in situations under consideration.} Proposition
\ref{I3.1} to define the angular velocities $\Omega_a$, and to deduce
the existence of a Killing vector $Y$ with periodic orbits in $\cDs$
when at least one of the $\Omega_a$'s is nonzero.  We have the
following preliminary result:
\begin{Proposition}
  \label{PP1}
Let $(M,g,X,\Sigma)$ satisfy the condition $\cC 1$. Then

1. There exists in $M$ an asymptotically flat maximal hypersurface
with boundary $\tilde \Sigma$, diffeomorphic to $\Sigma$, such
that\footnote{Actually $\cDs$ can be foliated by such surfaces, we
  shall, however, not need this result.}
$$
\prt \tilde \Sigma = \prt \Sigma, \qquad \cD(\tilde \Sigma) = \cDs\ .
$$

2.  $X$ is transverse to $\tilde \Sigma$; in particular all the
$\alpha_i$'s have the same sign and the gauge
condition\footnote{\label{fnote1}We do not assume here that the $U(1)$
bundle associated to the electro--magnetic field is trivial. Eq.\
\eq{new.1} should be viewed as a condition how to propagate some local
trivialization of the gauge bundle on $\tilde \Sigma$ to a
neighbourhood of $\tilde \Sigma$. When the gauge bundle is trivial,
then \eq{new.1} can be imposed globally because of the assumed
simple--connectedness of $\Sigma$.}
\begin{equation}
{\cal L}_X A_\mu = 0\
  \label{new.1}
\end{equation}
can be introduced\footnote{\label{Liefoot} Here the Lie derivative of
  $A$ is defined formally as that of a vector field.}, with
the (perhaps locally defined) potentials $A_\mu$ satisfying the
fall--off conditions \eq{spacetimefalloff3}, and being
continuous up--to--boundary on $ \overline{\Sigma}$.

3. If $X|_{\prt\Sigma}\ne 0$, then the canonical angular momentum
$J_i=J_i[Y,\tilde \Sigma]$ of each of the asymptotic three--ends
$\tilde\Sigma_i$ is well--defined and finite.  Here $Y$ is defined by
Proposition \ref{I3.1}.
\end{Proposition}
{\sc Proof:} Point 1 together with transversality of $X$ to $\tilde
\Sigma$ has been proved in \cite{ChWald}.   The existence of the
(perhaps local) gauge \eq{new.1}
follows from the fact that $X$ is transverse to $\tilde \Sigma$.
Note, however, that because $X$ is tangent to $\prt \Sigma$ the gauge
\eq{new.1} could become singular at $\prt \Sigma$. This is not the
case, and can be seen as follows: Near a connected component of
$\prt\Sigma$ one can introduce ``Rindler--type" coordinates adopted to
the action of the group of isometries generated by $X$, as in the
proof of Lemma 4.1 of \cite{ChWald}. In these coordinates one can
write a fairly explicity formula for a function $\lambda$ such that
$A_\mu + \prt_\mu\lambda$ satisfies \eq{new.1}, and the uniform
boundedness of the gauge potential in the new gauge readily follows.

To prove point 3, recall that the canonical angular momentum consists
of two parts ({\em cf.\ e.g.\/} \cite{ChAIHP} or \cite{SW1}), one
being the standard ADM angular momentum and the second coming from the
electro--magnetic field.  To take care of the ADM part, note that by
Proposition \ref{PAF1} in an appropriate coordinate system the fields
satisfy the fall--off conditions \eq{spacetimefalloff3}. Moreover
uniqueness results for maximal surfaces show that $Y$ must be tangent
to $\prt \Sigma$, which implies that
\begin{equation}
{\cal L}_X\gamma={\cal L}_XK=0\ .
\label{new.00}
\end{equation}
[Here $\gamma$ and $K$ are the induced metric and the extrinsic
curvature of $\tilde \Sigma$.]  The correctness of definition of the
ADM angular momentum follows from \eq{new.00} and from
\cite{Changmom}. To take care of the electro--magnetic contribution to
$J_i$, let $\phi_t[Y]$ be the one parameter group of diffeomorphisms
generated by $Y$. By assumption the (perhaps duality rotated) gauge
bundle is trivial on each asymptotic end $\tilde\Sigma_i$, and we can
choose $\prt\Sigma_i$ to be invariant under $\phi_t[Y]$. Let $\bar A$ be
any gauge--potential satisfying \eq{new.1} and the fall--off
conditions \eq{spacetimefalloff3}, define
\begin{equation}
A=\frac{1}{2\pi}\int_0^{2\pi} \phi_t[Y]^* \bar A dt\ .
  \label{newasdf}
\end{equation}
We have
\begin{eqnarray}
dA & = & \frac{1}{2\pi}\int_0^{2\pi} d\{\phi_t[Y]^* \bar A \}\, dt
=  \frac{1}{2\pi}\int_0^{2\pi} \phi_t[Y]^* d\bar A \, dt
\nonumber
\\
& = & \frac{1}{2\pi}\int_0^{2\pi} \phi_t[Y]^* F \, dt = F \ ,
\nonumber  \label{asdf}
\end{eqnarray}
so that $A$ is indeed a potential for $F$. Moreover we clearly have
$$
{\cal L}_Y A = 0,\qquad {\cal L}_X A = 0 \ ,
$$ the latter equation holding because $X$ and $Y$ commute. The
finiteness of the electromagnetic contribution to the canonical
angular momentum follows now from eq.\ (27) of \cite{SW2}.
\hfill$\Box$

\begin{Theorem}
\label{T3}
Let $(M,g,X,\Sigma)$ satisfy the condition $\cC 1$. Assume moreover
that the $U(1)$ bundle associated to the electromagnetic field can be
trivialized by performing a duality rotation and  that
\begin{equation}
\Omega_1=\ldots=\Omega_I=\Omega\ .
  \label{new.0}\end{equation}
Here $I$ is the number of connected components of $\partial \Sigma$
and the $\Omega_a$'s are their angular velocities, as defined by
Proposition \ref{I3.1}.
Then:

1. We necessarily have
\begin{equation}
\Omega(J_1+\cdots+J_K)\ge 0\ ,
  \label{new.0.1}
\end{equation}
where $K$ is the number of asymptotic ends of $\Sigma$, and the
$J_i$'s are the canonical angular momenta as defined in Proposition
\ref{PP1}.

2.  If the equality in \eq{new.0.1} is attained, then
$I=K=1$ and $\langle\langle \scri_\Sigma \rangle\rangle$ is isometrically
diffeomorphic to a connected component of the domain of outer
dependence of a (perhaps electrically and magnetically charged)
Reissner--Nordstr\"om black hole.
 \end{Theorem}

 \proof Proposition \ref{PP1} shows that the arguments of \cite{SW1}
 or \cite{SW2} apply. [The generalization of those arguments to the
 case in which several asymptotic ends are present, and in which
 $\prt\Sigma$ has several connected components but \eq{new.0} holds,
 presents no difficulties.] In particular when \eq{new.0.1} is
 actually an equality the staticity and the vanishing of the
 electromagnetic field follow from \cite{SW1,SW2}. Point 2 follows
 then from Ruback's uniqueness theorem \cite{Ruback}.  \hfill $\Box$

 It would be of interest to remove the condition \eq{new.0} above, the
 hypothesis of simple connectedness of $\Sigma$, as well as the
 condition of triviality\footnote{It appears that the hamiltonian
   formalism used in \cite{SW1} assumes at the outset the trivialiaty
   of the gauge bundle. This formalism can, however, be replaced by
   {\em e.g.\/} that of \cite{KT}, where no such restriction is
   needed.} of the $U(1)$ bundle associated to the electro--magnetic
 field.

\subsection{A uniqueness Theorem for space--times without black holes}
A well known theorem of Lichnerowicz \cite{Lichne} asserts that a {\em
  strictly} stationary vacuum space--time with a hypersurface
satisfying the conditions of Definition \ref{D1} and with {\em one}
asymptotically flat three--end is necessarily flat. Here {\em
  strictly} stationary is defined as the requirement that the Killing
vector $X$ approaches asymptotically the unit normal to $\Sigma$ and
is timelike everywhere. In this Section we shall present an extension
of this Theorem to the case where 1) many asymptotically flat ends are
potentially allowed, 2) the Killing vector is not {\em a priori}
assumed to be timelike everywhere, and 3) a potentially non--vanishing
electro--magnetic field is allowed. The proofs are mainly based on the
results of Sudarsky and Wald, and run very much in parallel with those
of the previous Section. The results in this Section are actually
rather more elegant, as one avoids all the technicalities related to
the bifurcation surfaces previously needed. Let us again
start with a Definition:

\begin{Definition}[Condition ${\cal C}2$]
  \label{DC2} A quadruple $(M,g,X,\Sigma)$ will be said to {\em
    satisfy the condition $\cC 2$} if $(M,g)$ is a maximal globally
    hyperbolic electro--vacuum space--time with electromagnetic field
    $F$ and if moreover the following conditions are satisfied:

    1. $\Sigma$ is a simply connected$^{\ref{simply}}$ spacelike
    hypersurface in $M$ satisfying the requirements of Definition
    \ref{D1}.

2. $X$ is a Killing defined on $\cD(\Sigma)$ such that ${\cal
        L}_X F=0$.  Moreover there exist constants $\alpha_i\in\R $ such
      that  on every asymptotic three--end $\Sigma_i$ of $\Sigma$ we
      have (after performing a Lorentz ``boost" of $\Sigma_i$ if
      necessary)
      \begin{equation}
        X\Big|_{\Sigma_i}\to_{r\to\infty}\alpha_i n\ ,
        \label{C2.1}
      \end{equation}
      where $n$ is the unit future directed normal to $\Sigma$. We
      shall normalize$^{\ref{samesign}}$ the $\alpha_i$'s so that $
      \alpha_1 = 1 $.

3. Let the domain of outer communication $ \langle\langle \scri_\Sigma
      \rangle\rangle$ be defined by \eq{doc}. We shall require that
      \begin{equation} \cD(\Sigma)\subset \langle\langle \scri_\Sigma
        \rangle\rangle\ .
\label{C2.3}
\end{equation}
In other words, there is no black hole or white hole in $\cDs$.
  \end{Definition}

  From \cite{ChWald} one has, in parallel to Proposition \ref{PP1},
  the following (the existence of the potentials  $V_i$ below follows
  from \eq{spacetimefalloff3} and from \cite{SW1}):
\begin{Proposition}
  \label{PP2}
Let $(M,g,X,\Sigma)$ satisfy the condition $\cC 2$. Then

1.  $M$ can be foliated by a family of asymptotically flat maximal
hypersurface (without boundary) $\tilde \Sigma_\tau$, $\tau\in\R$,
which are Cauchy surfaces for $\cDs$.

2.  $X$ is transverse to $\tilde \Sigma$; in particular all the
$\alpha_i$'s have the same sign and the gauge
condition$^{\ref{fnote1}}$
\begin{equation}
{\cal L}_X A_\mu = 0\
  \label{new.1.pps}
\end{equation}
can be introduced.$^{\ref{Liefoot}}$ In this gauge the limits
$$
V_i \equiv\lim_{r\to\infty}A_0\Big|_{\Sigma_i}
$$
exist and are angle--independent constants.
\end{Proposition}
The main result of this Section is the following:

\begin{Theorem}
\label{T4}
Let $(M,g,X,\Sigma)$ satisfy the condition $\cC 2$. Assume moreover
that the $U(1)$ bundle associated to the electromagnetic field can be
trivialized by performing a duality rotation.
Then:

1. The maximal surfaces $\Sigma_\tau$ foliating $\cDs$ ({\em cf.\/}
Proposition \ref{PP2}, Point 1) have vanishing extrinsic curvature,
and the pull-back of the electromagnetic field two--form $F$ to each
$\Sigma_\tau$ vanishes.

  2.  If moreover
\begin{equation}
\sum_{i=1}^K \alpha_iV_iQ_i = 0\ ,
  \label{new.2.1}
\end{equation}
then $(M,g)$ is the Minkowski space--time. Here $K$ is the number of
asymptotic three--ends of $\Sigma$, $V_i$ is the electrostatic
potential of the $i$'th end $\tilde\Sigma_i$ as defined in Proposition
\ref{PP2}, Point 2, $Q_i$ its total charge, and the $\alpha_i$'s are
the constants of Definition \ref{DC2}.
 \end{Theorem}

{\sc Remark:} \eq{new.2.1} necessarily holds when we have
$$
\alpha_1V_1=\ldots=\alpha_KV_K\ .
$$ Indeed, under the conditions of Definition \ref{DC2} the total
charge $\sum_iQ_i$ necessarily vanishes, and \eq{new.2.1} follows.

\proof Proposition \ref{PP2} allows one to apply the results of
\cite{SW1,SW2}, so that point 1 immediately follows. (2.10) and eq.\
(41) of \cite{SW2} (appropriately generalized to the case of a finite
number of asymptotic ends) show that $F\equiv 0$, so that $(\cDs,g)$
is vacuum.  As shown in \cite{Beig} ({\em cf.\/} also \cite{AAM}) in
each of the asymptotic ends the Komar integral of the Killing vector
$X$ is equal to $\alpha_im_i$, where $m_i$ is the ADM mass of the
$i$'th end. In vacuum the divergence of the Komar integrand
vanishes, so that we obtain
$$
\sum_{i=1}^K\alpha_i m_i =0\ .
$$ By Proposition \ref{PP2}, point 2, all the $\alpha_is$ have the
same sign, and the result follows from the positive energy theorem.
\hfill$\Box$

It would be desirable to remove condition \eq{new.2.1} above, the
hypothesis of simple connectedness of $\Sigma$, as well
as the hypothesis of the triviality of the $U(1)$ bundle associated to
the electromagnetic field.

\section{Folklore, Conjectures --  continued}
\label{continued}
\subsection{Rigidity and analyticity}
\label{analyticity}
The results discussed up to now allow one to obtain a reasonably
satisfactory version of Theorem \ref{T0} for non--rotating black
holes, {\em cf.\/} Theorem \ref{T3} above. To the list of problems
listed in Section \ref{introduction} we wish to add some further
problems
 which arise when considering the rotating black holes. The
key results concerning those are
\begin{enumerate}
  \item\label{rigid} Hawking's rigidity Theorem, and
  \item the Carter--Bunting--Mazur uniqueness Theorem.
\end{enumerate}
Recall that Hawking \cite{HE} has proved that the isometry group of an
analytic, electro--vacuum, stationary, non--static, asymptotically
flat space--time with a complete Killing vector $X$ and which contains
a black--hole must be at least two--dimensional ({\em cf.\/} \cite{HE}
for a precise description of the notions used). It has already been
pointed out by Carter \cite{CarterCargese} that the hypothesis of
analyticity here is rather unsatisfactory: Indeed, it is well known that
in regions where a Killing vector is timelike the metric must be
analytic (in appropriate coordinates) \cite{MzH}.  However, this needs
not to be true in those regions in which the Killing vector becomes
null or spacelike. As the Killing vector cannot be timelike on the
black--hole boundary, the hypothesis that the metric be analytic
up--to--and--including the even horizon made in \cite{HE} has no
justification. [Even in space--times without ergoregions, in which the
``stationary" Killing vector becomes null at the event horizon, the
metric needs {\em not} to be analytic {\em up to the horizon}. A
simple example illustrating the fact, that an analytic function needs
not to be analytic up to boundary is the following: Let $g$ be any
smooth real valued function defined on $\prt B(1)$, where $B(1)$
denotes the closed unit disc in $\R^2$, and suppose that $g$ is {\em
  not} real analytic.  Let $f:B(1)\to\R$ be a solution fo the equation
$\{(\partial/\partial x)^2 + (\partial/\partial y)^2\}f=0$ such that
$f|_{\prt B(1)}=g$. $f$ is real analytic on $\mbox{int}\,B(1)$ and
clearly {\em does not} extend to an analytic function on $B(1)$.]
Perhaps the most important open problem in the uniqueness theory of
black holes is therefore the following:

\begin{Problem}
\label{Panalycity}
Prove Hawking's rigidity Theorem {\em without assuming analyticity} of
$(M,g)$, or construct a counterexample.
\end{Problem}

\subsection{Isometry groups in asymptotically flat space--times}
Whatever the status of Hawking's rigidity for non--analytic
space--times, it is of interest to classify those stationary
asymptotically flat space--times which have more than one Killing
vector. Here, basing on what has been said above, one expects that the
solutions will be either flat, or spherically symmetric, or
axisymmetric. In other words, if there exists a Killing vector $X$
which is timelike in the asymptotically flat ends, then there will be
at least one more Killing vector field $Y$ which \begin{enumerate}
\item
has an axis of symmetry ({\em i.e.}, the set $\{p: Y(p)=0\}$ is
non--empty), and
\item
the orbits of which are periodic.
\end{enumerate}
A Killing vector satisfying the above will be called an {\em
  axial  Killing vector}. We believe that the following should
be true:

\begin{Conjecture}
\label{Lie}
  Let $(M,g)$ be a maximal globally hyperbolic electro--vacuum
  space--time with an 
  asymptotically flat 
  spacelike surface $\Sigma$ satisfying the requirements of Definition
  \ref{D1} or \ref{D2}, and with a Killing vector $X$ which is
  timelike in the asymptotically flat three--ends of $\Sigma$. Let
  $G_0$ be the connected component of the identity of the group of all
  isometries of $({\cal D}(\Sigma),g|_{{\cal D}(\Sigma)})$. Then
  \begin{enumerate}
  \item $G_0=\R\times SO(2)$, with axial  generator of the
    $SO(2)$ factor, or
  \item $G_0=\R\times SO(3)$, with two--dimensional spheres as
    principal orbits of the $SO(3)$ factor, or
      \item $G_0$ is the connected component of the identity of the
        Poincar\'e group.
  \end{enumerate}
  [Here the $\R$ action is that by time translations].
\end{Conjecture}

  In other words, if  $Y\ne X $ is a Killing vector field on $M$ which is {\em
not}
  axial , then the Killing Lie algebra of $(M,g)$ is that
of the Poincar\'e group.
Some 
results concerning this question can be found in
\cite{AX} ({\em cf.\/} also \cite{AshtekarSchmidt}) under, however,
some supplementary 
conditions. 

\subsection{Topology of black holes}
\label{topology}
To continue with our long list of problems in the uniqueness theory of
black holes, recall that the key to the Carter--Bunting--Mazur
uniqueness theorem for axisymmetric black holes is Carter's reduction
of the problem to a two--dimensional harmonic map boundary value problem
\cite{CarterLesHouches,CarterHI}. In that construction one assumes
that $(M,g)$ is asymptotically flat in the conformal sense, and that
on $M$ there exist two Killing vector fields, $X$, which approaches
$\prt/\prt t$ in asymptotically Minkowskian coordinates $(t,\vec x)$
as $r\to\infty$, and $Y$, which is an axial Killing
vector.\footnote{Let us mention that for spherical bifurcation
  surfaces the discussion of Section \ref{Wald} guarantees the
  existence of an axial Killing vector field in space--time, so that
  under the hypotheses of Proposition \ref{I3.1} the Carter reduction
  process can be applied.} One moreover assumes that the boundary of
the black hole is connected
and has spherical topology. Under these
assumptions Carter reduces the field equations to a two--dimensional
harmonic--map problem with appropriate boundary conditions
\cite{CarterHI}, which has subsequently been shown to have unique
solutions \cite{Mazur,Bunting,CarterCMP,Mazurreview}. In this context the
establishing of the validity of Conjecture \ref{Lie} would be rather
useful, reducing the general question of classification of stationary
asymptotically flat space--times with more then one Killing vector
field to that of axisymmetric black--holes considered by Carter.
Further improvements of the uniqueness theory of axisymmetric black
holes should include
\begin{enumerate}
\item
a justification of the black--hole--connectedness condition, and
\item
a justification of the spherical topology condition.
\end{enumerate}
Recall that a well--known claim of Hawking \cite{HE} asserts that a
connected component of a black hole boundary must necessarily have
spherical topology. The arguments used in \cite{HE} suffer from two
problems:
\begin{enumerate}
\item As has been emphasized by G.\ Galloway \cite{Galloway}, the
  claim in \cite{HE} that a black hole boundary cannot have toroidal topology
  does not seem to be sufficiently justified.\footnote{It seems that
    the arguments of \cite{Hawking} and \cite{HE} can be used to
    eliminate toroidal black holes when analyticity of the metric is
    assumed.} Moreover
\item
for degenerate black--holes one should  justify the degree of
differentiability of the black--hole boundary used in the proof.
\end{enumerate}
The new argument of \cite{Galloway} eliminates the toroidal black
holes at the price, however, of introducing 1) a condition on null
geodesics in $(M,g)$ and 2) a ``well--formedness'' condition on the
event horizon, {\em cf.\/} \cite{Galloway} for details.  It would be
of interest to perhaps justify\footnote{{\em cf.\ e.g.\/}
\cite{ChWald}[Proposition 3.5] for a result on the
``well--formedness'' condition.} those assumption for electro--vacuum
stationary space--times.

\section{Conclusions}

In this review we have attempted to present an exhaustive list of open
problems of the theory of uniqueness of black holes. A possible
approach to a more satisfactory theory is as follows:
The establishing of
Conjectures \ref{C1} and \ref{nonullconjecture} would
\begin{enumerate}
\item lead to a considerably improved version of Theorem \ref{T3}.
  Moreover
\item
Proposition \ref{I3.1} together with some perhaps improved version of
the topology--of--black--holes--theorem would lead to a complete
classification of stationary space--times with two or more Killing vectors.
\end{enumerate}
We believe that the establishing of Conjecture \ref{nonullconjecture},
and perhaps also of Conjecture \ref{Lie}, should be a
not--too--difficult Corollary of the positive energy
theorems.
A proof of Conjecture \ref{C1} would considerably improve our
knowledge of the structure of stationary asymptotically flat
space--times. Finally, a solution to Problem \ref{Panalycity} would be
major progress in mathematical general relativity.

\appendix
\section{Obstructions to smooothness of Scri for a class of
  stationary asymptotically Minkowskian space--times.}
\label{obstruction}

Let $M=\R^4\setminus\{\R\times B(R)\}$, where $B(R)\subset\R^3$ is a
closed coordinate ball of radius $R$, and let $g_{\mu\nu}$ be a metric
on $M$ satisfying \bea & \label{metric1}
g_{\mu\nu}-\eta_{\mu\nu}-\frac{h_{\mu\nu}({x^i\over
    r})}{r}=O(r^{-1-\epsilon})\ , & \\ & \label{metric2}
\prt_{i_1}\ldots\prt_{i_k}
\Big\{g_{\mu\nu}-\eta_{\mu\nu}-\frac{h_{\mu\nu}({x^i\over
    r})}{r}\Big\}=O(r^{-1-k-\epsilon})\ ,
 &
\eea
with some $k\ge 1$, some functions
$h_{\mu\nu}\in C^k(\R^3)$, and some $0<\epsilon \le 1$. Define
$$
Y^\mu_\pm \prt_\mu = \prt_t\pm \frac{x^i}{r}\prt_i\ ,
$$ and set \beaa & C^\mu_\pm({x^i\over r}) = \lim_{r\to\infty}
r^2\Gamma^\mu_{\nu\rho}Y^\nu_\pm Y^\rho_\pm \ , & \\ & c^\mu_\pm =
\frac{1}{4\pi}\int_{S^2}C^\mu_\pm\, d^2S\ ,  & \\ &
e_\pm = \frac{1}{4\pi}\int_{S^2}\sum_i {x^i\over r}C^i_\pm\, d^2S
\ , & \\ & D^\mu_\pm=C^\mu_\pm-c^\mu_\pm-e_\pm Y^\mu_\pm\
. & \eeaa
Here $\Gamma^\mu_{\nu\rho}$ are the Christoffel symbols of
$g_{\mu\nu}$, and $ d^2S = \sin\theta\,d\theta\,d\varphi$. We have the
following:
\begin{Proposition}
  \label{PO.1}
The $D^\mu_\pm$'s are geometric  invariants.
\end{Proposition}
{\sc Proof:} Consider two coordinate systems $\{x^\mu\}$, $\{y^\mu\}$
in which \eq{metric1}--\eq{metric2} hold, let us denote by
${C^\mu}_{\pm_,\{x^\alpha\}}$, respectively
${C^\mu}_{\pm,\{y^\alpha\}}$, the quantitites $C^\mu_\pm$ calculated
in the coordinate system $\{x^\mu\}$, respectively $\{y^\mu\}$;
similarly for ${D^\mu}_{\pm,\{y^\alpha\}}$, etc.
Now it follows\footnote{Due to the existence of a Killing vector the
  arguments of \cite{Chmass1,Chmass2} can be considerably simplified,
  {\em cf.\/} the remark after Corollary 1 in \cite{Chmass1}.} from
the results in \cite{Chmass1,Chmass2}  that there exists a Lorentz
matrix ${\Lambda^\mu}_\nu$, a constant vector $A^\mu$, and functions
$B^\mu\in C^{k+1}(\R^3)$
such that \bea
\label{OBS.5} & y^\mu- {\Lambda^\mu}_\nu x^\nu - A^\mu\log r -
B^\mu({x^i\over r})= O(r^{-\epsilon}) \ , & \\ \nonumber &
\prt_{\sigma_1}\ldots\prt_{\sigma_{k+1}}\Big\{y^\mu-{\Lambda^\mu}_\nu
x^\nu - A^\mu\log r -
B^\mu({x^i\over r}) \Big\} = O(r^{-\epsilon-k-1})\ . &
\eea Suppose first that ${\prt\over\prt x^0}={\prt\over\prt y^0}$,
hence ${\prt y^\mu\over \prt x^0}=\delta^\mu_0$. It follows that
${\Lambda^\mu}_\nu $ is a rotation matrix, and performing the inverse
rotation if necessary we may without loss of generality assume that
${\Lambda^\mu}_\nu=\delta^\mu_\nu$.  A calculation shows that
$$
{C^\mu}_{\pm,\{y^\alpha\}}- {C^\mu}_{\pm,\{x^\alpha\}} =-
\lim_{r\to\infty} r^2\frac{\prt^2y^\mu}{\prt r^2} = A^\mu
\ ,
$$
and for ${\prt\over\prt x^0}={\prt\over\prt y^0}$ the result follows.

Suppose, finally, that ${\prt\over\prt x^0}\ne{\prt\over\prt y^0}$. By
\eq{OBS.5} there exists a constant $\alpha$ such that $\alpha
{\prt\over\prt x^0}-{\prt\over\prt y^0}$ is an asymptotically spacelike
Killing vector. It then easily follows from \eq{metric1}--\eq{metric2}
that $g_{\mu\nu}$ must be flat. Going to coordinates $\hat x^\mu$ where the
metric
is explicitly flat we clearly have ${C^\mu}_{\pm,\{\hat
  x^\alpha\}}=0$, and ${D^\mu}_{\pm,\{  x^\alpha\}}={D^\mu}_{\pm,\{
  y^\alpha\}}=0$ follows from the calculation above.
\hfill $\Box$

Recall now that in \cite{BMS} the existence of a system of coordinates
$(\hat u,\hat r, \hat \theta,\hat\varphi)$ has been postulated such
that, if we set
$$
\hat t=\hat u  +\hat r
\ ,
$$ and if we define the Cartesian coordinates $\hat x^i$ in terms of the
spherical coordinates $(\hat r,\hat\theta,\hat\varphi)$ in the
standard way, then
\begin{enumerate}
\item the metric in the coordinates $\hat x^\mu$  satisfies
  \eq{metric1}--\eq{metric2}, and
\item the curves 
$\{x^\mu(s)\}=\{(s,s\hat x^i/r)\}$ are null geodesics.
\end{enumerate}
Coordinates satisfying the above will be called {\em retarded Bondi
  coordinates}. The {\em advanced Bondi
  coordinates} are defined by reversing the time--orientation above.

As discussed after the proof of Proposition \ref{PAF1}, such
coordinates can be constructed whenever a conformal completion in
lightlike directions satisfying appropriate requirements exists.
Conversely, existence of Bondi coordinates implies that of conformal
completions, {\em cf.\ e.g.\/} \cite{Penrose}.

The main result of this Appendix  is the following:

\begin{Theorem} \label{TOBS}
Consider a metric on $M=\R^4\setminus\{\R\times
  B(R)\}$ satisfying \eq{metric1}--\eq{metric2} with $k\ge 2$.

1. Retarded Bondi coordinates exist if and only if
\begin{equation}
\label{OBS.3}
D^\mu_+=0 \ .
\end{equation}

2. Advanced Bondi coordinates exist if and only if
\begin{equation}
\label{OBS.4}
D^\mu_-=0 \ .
\end{equation}
 \end{Theorem}

{\sc Remarks:}
1. Proposition \ref{PAF1} implies that \eq{OBS.3}--\eq{OBS.4} hold for
electro--vacuum stationary space--times.

2.  It is worthwile mentioning that \eq{OBS.3}--\eq{OBS.4} provide a
rather effective criterion. Consider, for example, a metric of the
form
\begin{equation}
  g_{\mu\nu}= -(1-\frac{2m(\theta,\varphi)}{r})dt^2+(1-\frac{2\tilde
    m(\theta,\varphi)}{r})^{-1}dr^2 +r^2 (d\theta^2 + \sin^2\theta\,
  d\varphi^2)\ ,
  \label{OBX}
\end{equation}
with some twice differentiable functions $m(\theta,\varphi)$ and
$\tilde m(\theta,\phi)$. One easily finds that the metric \eq{OBX}
admits Bondi coordinates only if $m$ and $\tilde m$ are constants,
with $m=\tilde m$.

3. If $g_{\mu\nu}$ admits a full expansion in terms of inverse powers
of $r$ for large $r$ and if \eq{OBS.3} holds, then the transformed
metric in the Bondi coordinates will also admit a full expansion.

4. If \eq{OBS.3}--\eq{OBS.4} do not hold, one can construct
``Bondi--type" coordinates in which the metric has $r^{-1}\ln r$
terms. If $g_{\mu\nu}$ admits a full expansion in terms of inverse
powers of $r$ for large $r$ but \eq{OBS.3} does not hold, or if the
metric admits an asymptotic expansion in terms of functions
$r^{-j}\ln^ir$, then the transformed metric in the ``Bondi--type"
coordinates will have an expansion in terms of $r^{-j}\ln^ir$.

5.  This result can essentially be found in \cite{Madore}; the
conclusions there are somewhat less definitive due to the dynamical
character of the metrics considered.

{\sc Proof:} We shall only prove point 1, point 2 follows by reversing
time--orientation. We claim that \eq{OBS.3} is necessary. Indeed, in
Bondi coordinates the curves $\{x^\mu(s)\}=\{(s,s\hat x^i/\hat r)\}$ are null
geodesics so that we have
\begin{equation}
\frac{D^2x^\mu}{ds^2} = \Gamma^\mu_{\nu\rho}\frac{dx^\nu}{ds}
\frac{dx^\rho}{ds} \sim \frac{dx^\mu}{ds}= Y^\mu_+ 
\ .
  \label{OBS.6}
\end{equation}
\eq{OBS.6} shows that \eq{OBS.3} holds in Bondi's coordinates, hence
in any coordinate system by Proposition \ref{PO.1}. The sufficiency of
\eq{OBS.3} can be proved by constructing explicitly the appropriate
family of null geodesics. One  shows that if \eq{OBS.3} holds,
then the transformation leading to Bondi coordinates contains $\log r$
terms only in the form \eq{OBS.5}; no details will be given.
\hfill $\Box$

\section{Majumdar--Papapetrou space--times with an infinite number of
  black holes} \label{Veron} In this Appendix we shall briefly discuss
a class of Majumdar--Papapetrou space--times with an infinite number
of black holes considered by L.\ V\'eron\footnote{L.\ V\'eron, private
  communication.}.  Let $\{\vec a_i\}_{i=1}^\infty$ be an arbitrary
sequence of pairwise distinct vectors in $\R^3$, and let $m_i$ be an
arbitrary sequence of non--negative numbers satisfying
\begin{equation}
m\equiv \sum_{i=1}^\infty m_i < \infty\ .
\label{V.1}
\end{equation}
Consider the manifold $M=\R\times(\R^3\setminus \overline{\{\vec
  a_i\}
})$ with a metric of the form
  \begin{eqnarray}&
ds^2=-U^{-2}dt^2+U^2(dx^2+dy^2+dz^2)\ ,
    \label{V.2}
& \\ &
U = 1 + \sum_{i=1}^\infty \frac{m_i}{|\vec x - \vec a_i|}\ .
    \label{V.3} &
  \end{eqnarray}
  It is easily seen, using {\em e.g.\/} the Harnack principle and
  \eq{V.1}, that $U$ is a well defined smooth function on
  $\R^3\setminus \overline{\{\vec a_i\} }$ satisfying
  \begin{equation}
\Delta U = 0 \ .
    \label{V.4}
  \end{equation}
As has been shown in \cite{Majumdar,Papapetrou}, 
                                                 the metric \eq{V.2}
                                                 solves the
                                                 electro--vacuum
                                                 Einstein equations if
                                                 we set
\begin{equation}
A_\mu dx^\mu=U^{-1}dt\ .
  \label{V.5}
\end{equation}
Let us moreover assume that there exists $R>0$ such that $\{\vec
a_i\}_{i=1}^\infty$ is included in a ball of radius $R$.
By definition of $U$ and by known  properties of 
solutions of the Laplace equation  it can be seen that we have
$$  U-1-\frac{m}{r}=O(r^{-2}) \ ,
$$ with $m$ defined in \eq{V.1}, and with appropriately faster decay
of all the derivatives of $U-1-\frac{m}{r}$. It follows that $M$
contains an asymptotically flat four--end $M_1$ in the sense of
Section \ref{asymptotic}. The arguments of Section II of
\cite{HartleHawking} show that every ``point" $\vec x = \vec a_i$
such that $\vec a_i$ is not an accumulation point of $\{\vec
a_i\}_{i=1}^\infty$
corresponds to a connected component of the boundary of a black--hole
when the space--time is suitably analytically extended, with a
degenerate event horizon which has spacelike cross--sections of area
$4\pi m_i^2$. Thus, if there is an infinite number of non--vanishing
coefficients $m_i$ (which we shall henceforth assume), then the
resulting maximally analytically extended space--time contains a black
hole region (with respect to the asymptotic end $M_1$) with an
infinite number of connected components. Let us, however, note the
following:
\begin{enumerate}
\item The metric \eq{V.2} is ``nakedly singular", which can be seen
  as follows: By well known properties of solutions of Laplace
  equation for any point $\vec a_i$ which is not an accumulation point of
  $\{\vec a_i\}_{i=1}^\infty$ we have
$$
\lim_{\vec x \to\vec a_i}F_{\mu\nu}F^{\mu\nu} = \lim_{\vec x \to\vec
  a_i} \frac{|\mbox{grad}\, U |^2}{U^4} = \frac{1}{m_i^2}\ ,
$$ and since $m_i\to 0$ as $i$ tends to infinity the scalar
$F_{\mu\nu}F^{\mu\nu}$ is unbounded on any hypersurface
$t=\mbox{const}$. It is then easily seen that one can construct a
causal curve which reaches future null infinity and on which
$F_{\mu\nu}F^{\mu\nu}$ is unbounded.
\item We believe that the partial Cauchy surfaces $\Sigma_\tau\equiv\{t=\tau\}$
are
  not complete with respect to the induced metric in general. One
can, however, find solutions with an infinite number of black holes
for which the $\Sigma_\tau$'s will be complete.  This is {\em e.g.\/}
the case when the sequence $\{\vec a_i\}_{i=1}^\infty$ has $a_1$ as
the only  accumulation point.
\end{enumerate}
It has been suggested \cite{HartleHawking}[Section III] that the only
Majumdar--Papapetrou space--times without naked singularities
\footnote{When analytically extended through a past event horizon, the
  black hole solutions of \cite{HartleHawking} possess naked
  singularities in $J^-(\Sigma)$.} in $J^+(\Sigma)$, where $\Sigma$ is
the hypersurface $\{t=0\}$, are those with a $U$ of the form \eq{V.3},
with only a finite number of non--vanishing $m_i$'s. It would be of
some interest to prove such a result.

\medskip
\begin{center}
  {\bf Acknowledgements}
\end{center}
 Most of the work on this paper was done when
the author was visiting the Max Planck Institut f\"ur Astrophysik in
Garching; he is grateful to J\"urgen Ehlers and to the members of the
Garching relativity group for hospitality.
Special thanks are due to Robert Wald for many discussions, and for
pointing out a mistake in the previous version of this paper.

\end{document}